\documentclass[aps,prb,twocolumn,groupedaddress,longbibliography]{revtex4-1}

\usepackage{amsmath}
\usepackage{amssymb}
\usepackage{graphicx}
\usepackage[colorlinks=true,
pdfborder={0 0 0},
linkcolor=blue,
citecolor=black]{hyperref}

\begin{document}

\title{Circuit-Model Analysis for Spintronic Devices with Chiral Molecules as Spin Injectors}

\author{Xu~Yang}
\email[]{xu.yang@rug.nl}

\affiliation{Zernike Institute for Advanced Materials, University of Groningen, NL-9747AG Groningen, The Netherlands}

\author{Tom~Bosma}
\affiliation{Zernike Institute for Advanced Materials, University of Groningen, NL-9747AG Groningen, The Netherlands}

\author{Bart~J.~van~Wees}
\affiliation{Zernike Institute for Advanced Materials, University of Groningen, NL-9747AG Groningen, The Netherlands}

\author{Caspar~H.~van~der~Wal}
\affiliation{Zernike Institute for Advanced Materials, University of Groningen, NL-9747AG Groningen, The Netherlands}

\date{\today}

\begin{abstract}

Recent research discovered that charge transfer processes in chiral molecules can be spin selective and named the effect chiral-induced spin selectivity (CISS). Follow-up work studied hybrid spintronic devices with conventional electronic materials and chiral (bio)molecules. However, a theoretical foundation for the CISS effect is still in development and the spintronic signals were not evaluated quantitatively. We present a circuit-model approach that can provide quantitative evaluations. Our analysis assumes the scheme of a recent experiment that used photosystem~I (PSI) as spin injectors, for which we find that the experimentally observed signals are, under any reasonable assumptions on relevant PSI time scales, too high to be fully due to the CISS effect. We also show that the CISS effect can in principle be detected using the same type of solid-state device, and by replacing silver with graphene, the signals due to spin generation can be enlarged four orders of magnitude. Our approach thus provides a generic framework for analyzing this type of experiments and advancing the understanding of the CISS effect.

\end{abstract}

\maketitle

Electronic spin lies at the heart of spintronics due to its capability to convey digital information. In contrast, this quantum mechanical concept has found few applications in chemistry and biology as the energy states associated with opposite spin orientations are often degenerate. Molecular chirality, on the other hand, is thoroughly discussed in chemistry and biology but rarely concerned in spintronics. In the past decade, the two concepts have been increasingly linked together thanks to the discovery of the chiral-induced spin selectivity (CISS) effect, which describes that the electron transfer in chiral molecules is spin dependent.\cite{ray1999asymmetric,yeganeh2009chiral,gohler2011spin,xie2011spin,guo2012spin,gutierrez2012spin,medina2015continuum,michaeli2016electron,matityahu2016spin,naaman2012chiral,naaman2015spintronics} This discovery not only provides new approaches to controlling chiral molecules~\cite{banerjee2018separation} and understanding their interactions,\cite{naaman2018chirality} but also opens up the possibility of small, flexible, and fully organic spintronic devices. Previously, organic materials were incorporated in spintronic devices as spin transport channels and spin-charge converters, but the conversion efficiency remained low.\cite{naber2007organic,dediu2009spin,sun2014first,dediu2002room,xiong2004giant,fang2011electrical,ando2013solution,sun2016inverse,liu2018organic} Building on CISS, hybrid devices with efficient molecular spin injectors and detectors were realized.\cite{kumar2013device,dor2013chiral,mathew2014non,carmeli2014spin,peer2015nanoscale,mondal2015chiral,eckshtain2016cold,koplovitz2017magnetic,michaeli2017new,varade2017bacteriorhodopsin} However, a full understanding of the signals produced by these devices is still lacking, and thereby the understanding of CISS largely hindered. 

We present here a circuit-model approach to quantitatively evaluating the spin signals measured from hybrid solid-state devices designed for studying the CISS effect. Similar approaches have been used for the analyses of spintronic devices with metallic and semiconducting materials.~\cite{banhart1997applicability,brataas1999spin,jedema2003spin} They provided accurate descriptions of experimental results and have been extended to a wide range of device geometries. We apply here such modeling to devices with adsorbed molecular active layers instead of metal contacts. While generally applicable, we take the device reported in Ref.~\onlinecite{carmeli2014spin} as a case study for demonstrating our approach. In comparison to our recent analysis using electron-transmission modeling,\cite{yang2019spin} the circuit-model approach is more suited for including a role for optically driven chiral molecules, and for electron transport outside the linear-response regime.

\begin{figure}[hbt]
	
	\includegraphics[width=\linewidth]{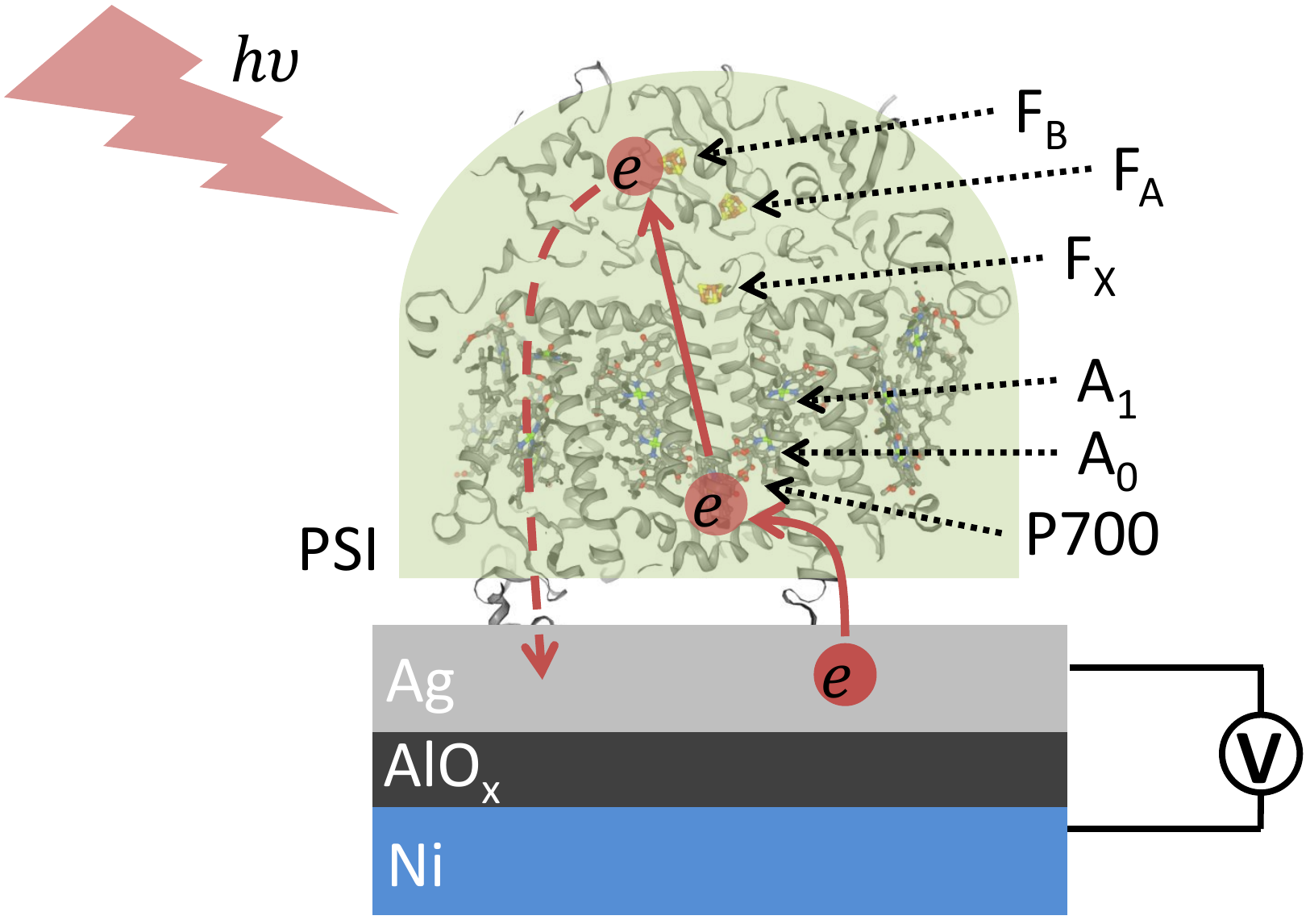}
	\caption{Electron transfer chain of PSI and the geometry of the solid-state device used in Ref.~\onlinecite{carmeli2014spin}. The device was a stack of 150~nm of nickel, 0.5~nm of AlO$_\text{x}$, and 50~nm of silver. PSI was immobilized on top of the silver layer, and the voltage difference between the silver and the nickel was measured. PSI is represented by the green area, on which the structure of a part that contains the PSI electron transfer chain is overlaid. The structure highlights key cofactors such as Fe$_{\text{4}}$S$_{\text{4}}$ clusters (F$_{\text{B}}$, F$_{\text{A}}$ and F$_{\text{X}}$), primary electron acceptors (A$_1$ and A$_0$), the reaction center (P700), and the chiral (helical) structural surroundings. Here PSI is in the \textit{up} orientation, with P700 close to silver, and the Fe$_{\text{4}}$S$_{\text{4}}$ clusters at the far end. Red labellings mark the light induced electron transfer process, including the photon ($h\nu$) and the electron ($e$), the photo-excitation pathway (solid arrows) and the unknown relaxation pathway (dashed arrow). The protein structure is taken from the RCSB Protein Data Bank (PDB ID 1JB0).\cite{jordan2001three}}
	\label{fig:device}
	
\end{figure}

\begin{figure*}[htb!]
	\includegraphics[width=\linewidth]{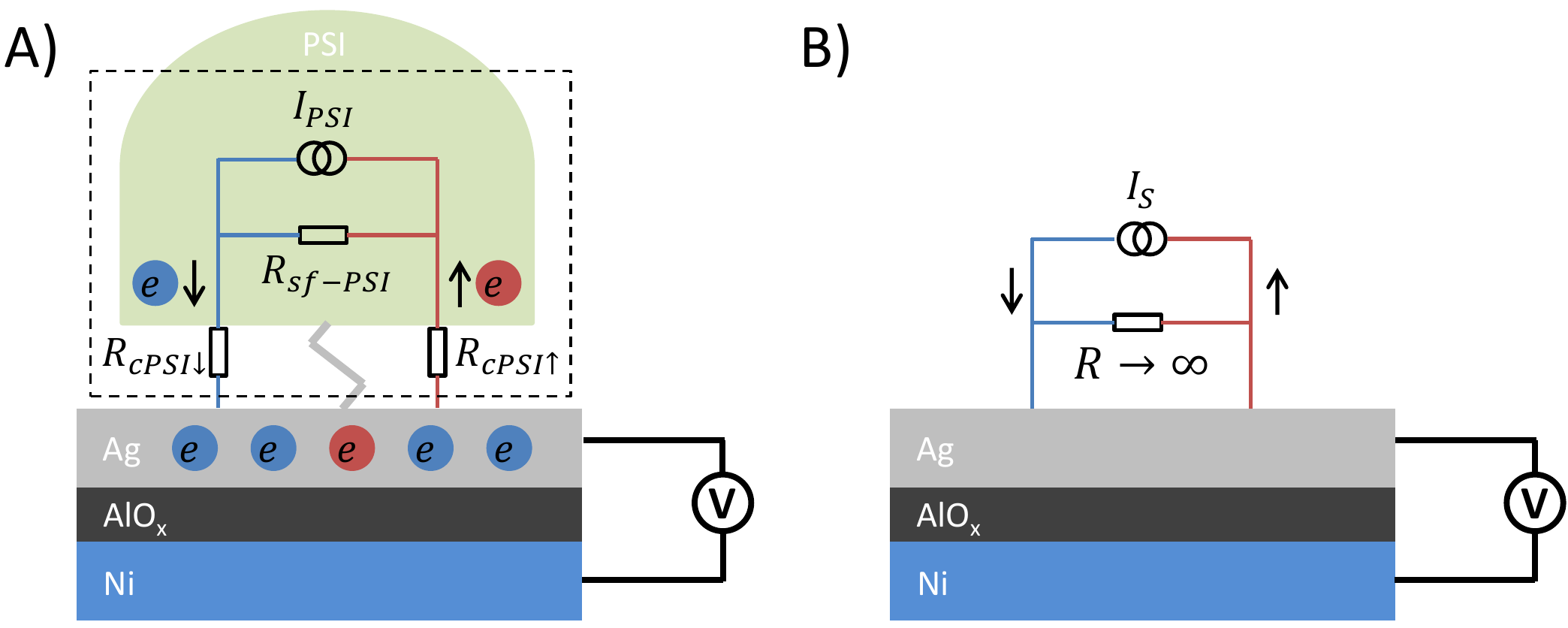}  
	\caption{PSI modeled as a spin-current source. \textbf{A)} PSI compared to a pure spin-current source with a spin relaxation pathway. It creates spin accumulation in the silver layer upon light illumination. Spin-up electrons (red) are transferred from silver to PSI, while spin-down electrons (blue) are transferred back to silver. No charge current flows through PSI but a spin-down accumulation is created in silver. \textbf{B)} The model of panel \textbf{A)}, reduced to its net spin-injection effect. The contribution from each component in the dashed box in \textbf{A)} cannot be clearly distinguished, therefore we treat them together as an ideal spin-current source $I_s$ with a parallel resistance $R$. Since the total impedance in PSI is much larger than that of silver, we consider $R\to\infty$. The net effect of this reduced model is to inject a spin current $I_s$ into the silver layer. Here we show the drawing for one PSI unit, but the spin currents ($I_{PSI}$, $I_s$) concern the values for the entire PSI ensemble on the device.}\label{fig:psi}
\end{figure*}

In the work of Ref.~\onlinecite{carmeli2014spin}, cyanobacterial photosystem~I (PSI) protein complexes were self-assembled on a silver-AlO$_\text{x}$-nickel junction and the orientation of PSI (\textit{up} or \textit{down}) was controlled by mutations and linker molecules. Figure~\ref{fig:device} shows a device with PSI in the \textit{up} orientation. Here, P700, the reaction center of PSI, was located adjacent to the silver layer. In P700, charge separation took place upon the illumination of a 660-nm laser during the experiments. It was described that the excited electron got transferred to the Fe$_{\text{4}}$S$_{\text{4}}$ clusters at the other end of PSI, and the hole left behind in P700 was refilled by an electron from silver. This process causes a net \textit{upward} electron transfer from silver to PSI, which, before relaxation, results in a steady-state increase of the silver surface potential, as was observed using a Kelvin probe.\cite{carmeli2014spin} In contrast, a device with PSI in the \textit{down} orientation gave a decrease of silver surface potential upon light illumination, indicating a net \textit{downward} electron transfer from PSI into silver. Both devices were then placed under laser illumination in the presence of an out-of-plane magnetic field which was used to set the magnetization of nickel in either the \textit{up} or \textit{down} direction. The charge voltage between silver and the nickel layer was monitored. The absolute value of this voltage was found to be always lower when the electron transfer direction and the magnetic field direction were parallel (both \textit{up} or both \textit{down}), and higher when they were anti-parallel (one \textit{up} and one \textit{down}). This magnetic field dependence suggested that the electron transfer process in PSI was spin selective, and the preferred spin orientation was parallel to the electron momentum. As PSI is one of Nature's two major light-harvesting centers, this intriguing result indicated that electron spins may also play a role in photosynthesis.	

However, an important question to address while considering this conclusion is: How much of the observed magnetic-field-dependent signal was from CISS? To answer this question we need to understand the origin of the measured steady-state magnetic-field-dependent voltage. Upon photo-excitation charge carriers were transferred from silver to PSI. These carriers must relax back to silver via pathways inside PSI because there was no top electrode providing alternative pathways. Both the excitation and relaxation pathways might exhibit spin selectivity. Qualitatively, as long as the CISS effects in the two pathways do not cancel each other, a net spin injection into silver can be generated. This spin injection then competes with the spin relaxation process in silver, and results in a steady-state spin accumulation which can indeed be detected as a charge voltage between silver and the nickel layer.\cite{jedema2002electrical}

To quantitatively evaluate this voltage signal, we adopt a two-current circuit model where spin transport is described by two parallel channels (spin-up and spin-down channels).\cite{mott1958theory,fert1968two} The two channels are connected via a spin-flip resistance $R_{sf}$, which characterizes the spin relaxation process in a nonmagnetic material. A derivation of $R_{sf}$ and a more detailed introduction of the two-current model concept can be found in Appendix~\ref{app:2Cmodel}. For a thin-film nonmagnetic material, we find
\begin{equation}\label{eqn:rsfag}
R_{sf} = 2 \cdot  \frac{\lambda_{sf}^2}{d \; A_{rel} \;\sigma}
\end{equation}
(assuming $d<\lambda_{sf}$ and $A_{rel} \gg \lambda_{sf}^2$),
where $\lambda_{sf}$ is the spin-relaxation length of the material, $\sigma$ is the conductivity of the material, $d$ is the thickness of the film, and $A_{rel}$ is the relevant area of the film where spin injection occurs. Notably, $R_{sf}$ is entirely determined by the properties of the material and the geometry of the device.

The role of PSI in the device can be characterized by two features. Firstly, due to the lack of a top electrode, there was (as a steady-state average) no net charge current flowing through PSI. Secondly, facilitated by CISS, PSI gave a net spin injection into silver. These two features resemble a pure spin-current source. Therefore, we model PSI as a pure spin-current source between the fully polarized spin-up (red) and spin-down (blue) channels, as shown in Figure~\ref{fig:psi}\textbf{A)}. Upon photo-excitation PSI sources an internal spin current $I_{PSI}$. The pathway with spin-flip resistance $R_{sf-PSI}$ accounts for the spin relaxation inside PSI. At the PSI-silver interface the two channels encounter possibly spin-dependent contact resistances $R_{cPSI\uparrow}$ and $R_{cPSI\downarrow}$. The net spin current injected from PSI into silver is $I_s=\eta \cdot I_{PSI}, (-1\leqslant \eta \leqslant 1)$, with $\eta$ being the fraction of the photo-induced spin current that actually contributes to the spin accumulation in silver. Generically, we regard PSI as a black box: a two-terminal unit that drives a spin current $I_s$, as shown in Figure~\ref{fig:psi}\textbf{B)}. This will later be linked and compared to known timescales for charge transfer processes inside PSI.

\begin{figure}[htb!]
	\includegraphics[width=\linewidth]{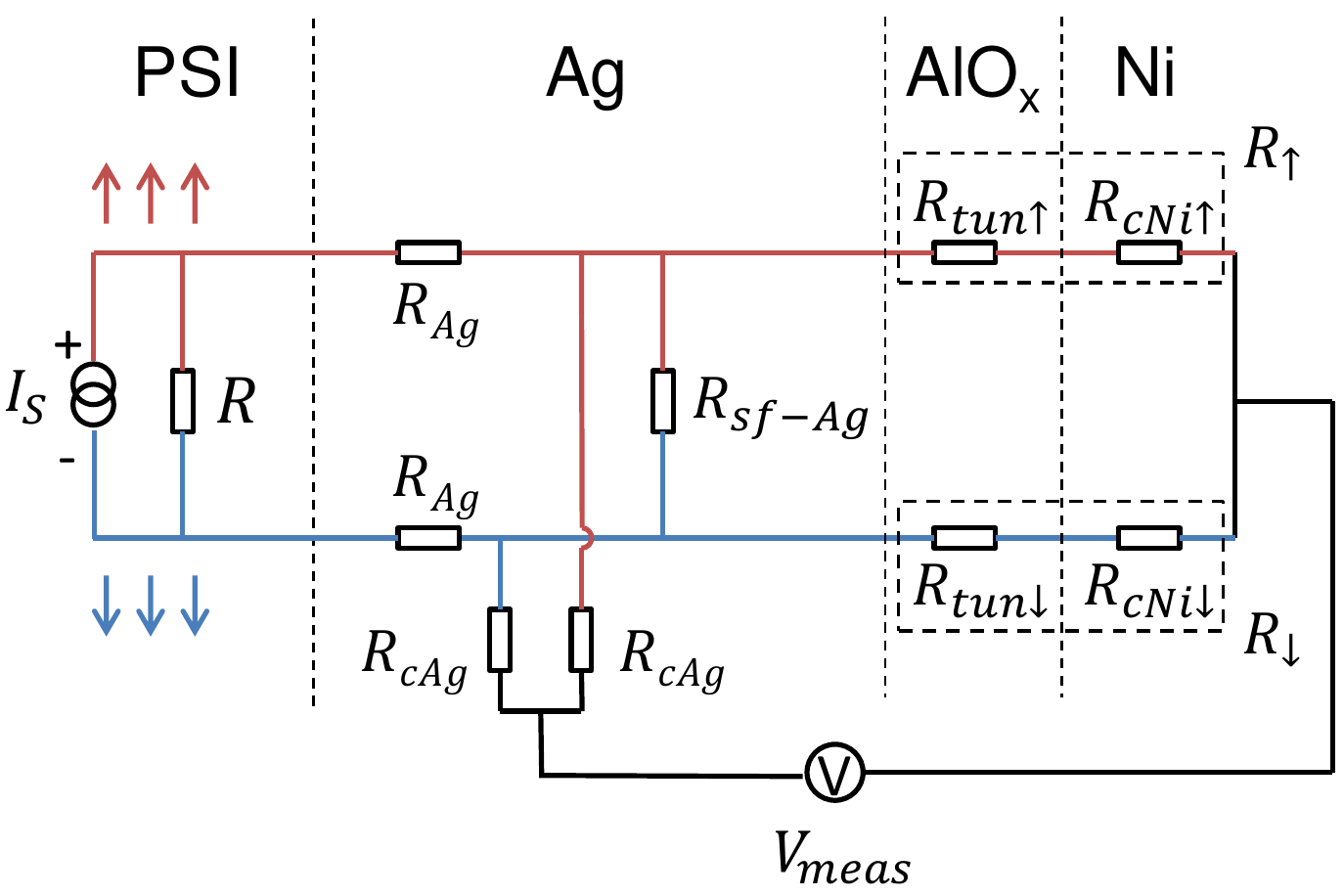}
	\caption{Two-current circuit model for the spintronic device of Ref.~\onlinecite{carmeli2014spin} (symbols introduced in the main text). Different parts of the device are separated by dashed lines. Spin-up and spin-down current channels are distinguished by color. PSI is represented by a pure spin-current source as introduced in Figure~\ref{fig:psi}\textbf{B)}. The spin relaxation in silver is modeled as a pathway with spin-flip resistance $R_{sf-Ag}$ connecting the two spin-current channels.}\label{fig:circuit}
\end{figure}

A circuit model for the entire device is shown in Figure~\ref{fig:circuit}. $R_{Ag}$ is the spin-independent resistance (in the out-of-plane direction) of the silver layer. Inside the silver layer the spins can relax, as represented by a spin-flip pathway with resistance $R_{sf-Ag}$. $R_{cAg}$ is the contact resistance between silver and the voltage meter. In principle these contacts could provide an extra pathway for electron spins to relax, but in reality these contacts are located millimeters away from where spins are injected. This distance is much larger than the spin-relaxation length in silver (about $150 ~\text{nm}$ at room temperature).\cite{godfrey2006spin} Therefore, the spin relaxation through these contacts is negligible and we can assume $R_{cAg} \to \infty$.

Underneath the silver layer is the the AlO$_\text{x}$ tunnel barrier and the ferromagnetic nickel layer. In these layers electrons experience spin-dependent resistances: the tunnel resistance $R_{tun\uparrow(\downarrow)}$ and the contact resistance $R_{cNi\uparrow(\downarrow)}$ (which includes the out-of-plane resistance of the nickel layer). Note that here the subscript $\uparrow$$(\downarrow)$ refers to the corresponding spin-current channel, not to be confused with the magnetization direction of nickel which determines the values of $R_{tun\uparrow(\downarrow)}$ and $R_{cNi\uparrow(\downarrow)}$. These resistances can be combined using shorter notations $R_{\uparrow} = R_{tun\uparrow} + R_{cNi\uparrow}$ and $R_{\downarrow} = R_{tun\downarrow} + R_{cNi\downarrow}$. An interchange of the $R_{\uparrow}$ and $R_{\downarrow}$ values thus accounts for the reversal of the magnetization direction of nickel.

The magnetization direction of nickel can be described as being parallel $(p)$ or anti-parallel $(ap)$ to the spin-up channel. For each case the reading of the voltage meter $V_{meas}$ is
\begin{subequations}
	\begin{equation}
	V_{meas}^{(p)} = \frac{1}{2} I_s (R_{\uparrow}-R_{\downarrow}) \frac{R_{sf-Ag}}{R_{\uparrow} + R_{\downarrow} + R_{sf-Ag}}~,
	\end{equation}
	\begin{equation}
	V_{meas}^{(ap)} = \frac{1}{2} I_s (R_{\downarrow}-R_{\uparrow}) \frac{R_{sf-Ag}}{R_{\uparrow} + R_{\downarrow} + R_{sf-Ag}}~.
	\end{equation}
\end{subequations}
The change in the measured voltage upon the reversal of the nickel magnetization is therefore
\begin{equation} \label{eqn:vdiff}
\begin{split}
V_{diff} &=V_{meas}^{(ap)} - V_{meas}^{(p)} \\
 &= I_s (R_{\downarrow}-R_{\uparrow}) \frac{R_{sf-Ag}}{R_{\uparrow} + R_{\downarrow} + R_{sf-Ag}} \\
 &= I_s R_{eff}~,
\end{split}
\end{equation}
where $R_{eff}=V_{diff}/I_s$ is an effective spinvalve resistance.

For the envisioned spintronic behavior in Ref.~\onlinecite{carmeli2014spin}, this model captures all relevant aspects for spintronic signals in the linear transport regime, without making assumptions that restrict its validity. It is thus suited for describing the observed spin signals in a quantitative manner when the values of the circuit parameters are available. For the device described in Ref.~\onlinecite{carmeli2014spin} we derive (see Appendix~\ref{app:Reff})
\begin{equation}\label{eqn:reff-est}
R_{eff} \approx 15 ~\text{m}\Omega~.
\end{equation}
Note that $R_{eff}$ is fully determined by the properties of the Ag-AlO$_\text{x}$-Ni multilayer, and deriving its value does not use any estimates or assumptions concerning PSI. Furthermore, by carefully choosing material parameters, the estimate of $R_{eff}$ is of great accuracy. This is also discussed in Appendix~\ref{app:Reff}.

This result for $R_{eff}$ directly yields values for the injected spin current that was flowing in the experiment of Ref.~\onlinecite{carmeli2014spin}.
For the \textit{up} orientation of PSI, the measured voltage difference $V_{diff}$ was about 50~nV. Thus, the net spin current injected into silver must have been $I_s=V_{diff}/R_{eff}\approx3~\mu\text{A}$. For the opposite PSI orientation, the measured $V_{diff}$ was about 10~nV, and accordingly, $I_s \approx 0.6~\mu\text{A}$.

Next, we turn these spin-current values into values for the timescale $\tau$ that must then hold for the charge excitation-relaxation process for illuminated PSI. Here $\tau$ can be understood as the time interval between two consecutive photo-excitation processes from the same PSI unit. By assuming that the intensity of the illumination is strong enough to drive all the PSI units in continuous excitation-relaxation cycles (saturated), we can write $i= -e/\tau$, where $e$ is the elementary charge and $i$ the photo-induced charge current in a PSI unit. The sum of all contributions $i$ (sum over all PSI units) should then be high enough to provide the above $I_s$ values. In order to check this, we will assume the highest number for PSI units that can contribute, and that they all maximally contribute. Therefore, we first assume that over the relevant area of the device the PSI units form a densely packed, fully oriented monolayer, and that all PSI units function identically. Secondly, we assume that photo-induced spin current from each PSI unit is fully injected into the silver layer, \textit{i.e.} $\eta=1$. Further, we assume that the polarization of the CISS effect in PSI is $50\%$, on par with the reported CISS polarization in other chiral systems.\cite{gohler2011spin,carmeli2014spin,michaeli2016electron} For these assumptions we find (details in Appendix~\ref{app:times}) that for the \textit{up} orientation of PSI, $\tau$ should not be larger than 100~ps. For the \textit{down} orientation this limit is $\tau \leqslant500~\text{ps}$. Note that the boundaries here correspond to the most ideal scenario, and in practice the required $\tau$ values could be much smaller than these boundaries.

\begin{figure}[hbt!]
	\includegraphics[width=0.9\linewidth]{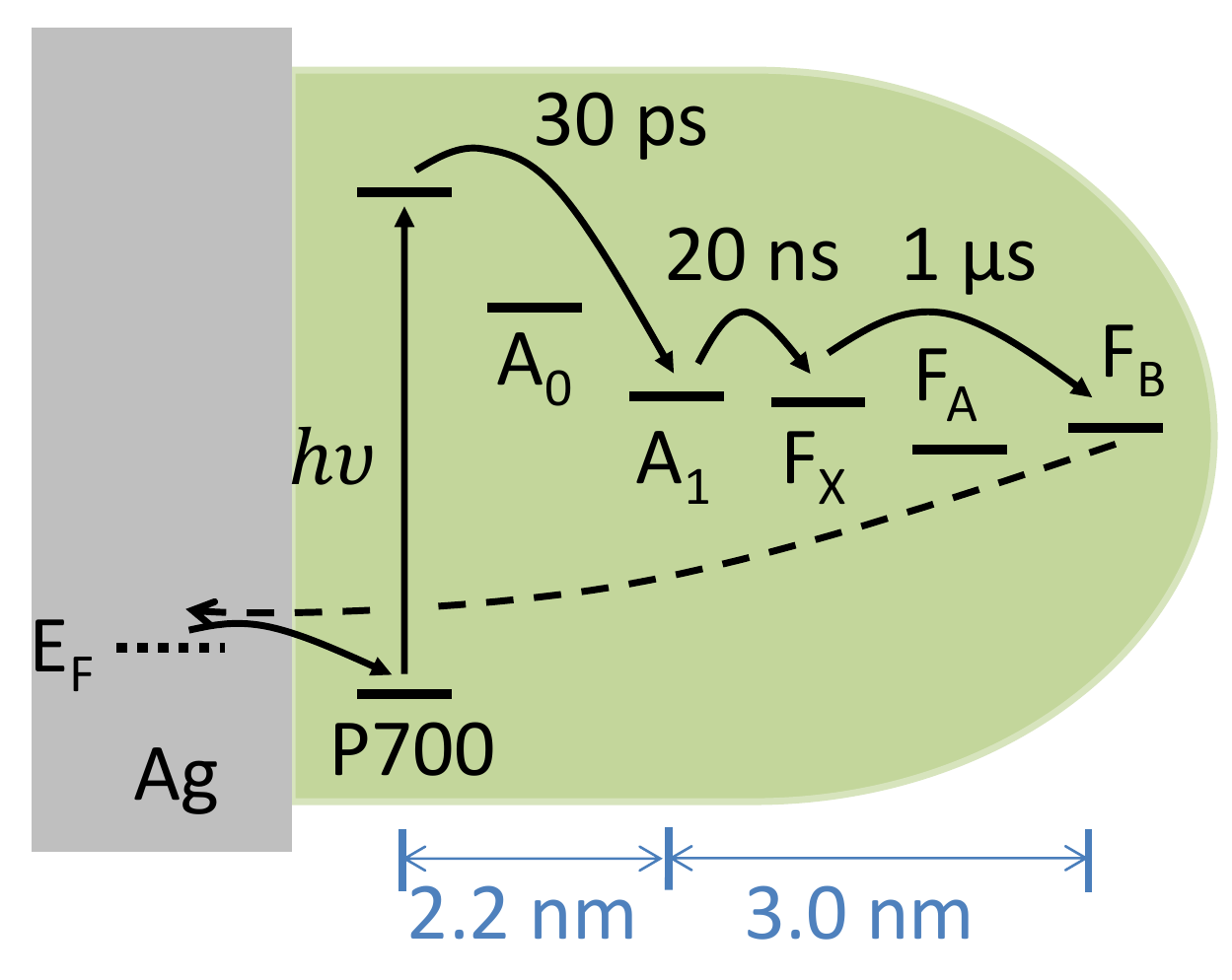}
	\caption{Electron transfer process and corresponding time scales in PSI, here depicted in a manner where the vertical placements of states reflect their energy levels. Here the PSI unit is in the \textit{up} orientation. The excitation process is labeled by the black arrows with corresponding time scales marked. However, the relaxation process is unknown (dashed arrow). The blue scale shows the spatial distance between different parts of the electron transfer chain.}\label{fig:times}
\end{figure}

We now compare these requirements for $\tau$ with the well-studied timescales of the electron transfer process in PSI. During photosynthesis, the photo-induced charge separation in PSI takes place at the primary donor P700. Electrons are then transferred through a series of accepters along the electron transfer chain: A$_{\text{0}}$, A$_{\text{1}}$, and the Fe$_{\text{4}}$S$_{\text{4}}$ clusters F$_{\text{X}}$, F$_{\text{A}}$ and F$_{\text{B}}$ (see Figure~\ref{fig:times}).\cite{jordan2001three,brettel1997electron,brettel2001electron} The initial electron transfer from P700 to $A_{\text{1}}$ is ultrafast ($\sim$30~ps), and further transfer to F$_{\text{X}}$ happens in 20-200~ns. Then, the electron transfer from F$_{\text{X}}$ through F$_{\text{A}}$ to F$_{\text{B}}$ typically takes 500~ns to 1~$\mu$s.\cite{brettel2001electron}

The requirements for $\tau$ values that we found are--regardless the PSI orientation--only compatible with the initial ultrafast electron transfer from P700 to $A_{\text{1}}$. The subsequent steps are at least two orders of magnitude too slow. Thus, concluding that the observed signals fully result from the CISS effect requires the existence of an ultrafast relaxation process where electrons immediately return to P700 after their initial transfer from P700 to $A_{\text{1}}$. 
This process does not exist in Nature, because it would stop the trans-membrane electron transfer in photosynthesis. We should, nevertheless, consider whether it can occur in the device, since PSI is there located in a very different environment. 

In the solid-state environment, faster relaxation than in Nature could be due to, for instance, the use of linker molecules, the mutations of PSI, or the presence of silver (thanks to its high density of states). The linker molecules are unlikely to be the reason, because their size is significantly smaller than PSI, and the electron transfer chain is positioned deeply in the center of PSI (Figure~\ref{fig:times}). Moreover, it was stated in Ref.~\onlinecite{carmeli2014spin} that the observed signals do not depend on the linker molecules. 
	
The mutations and the metal substrate, on the other hand, could indeed affect the electron transfer. To assess the effects, we can draw direct comparisons between Ref.~\onlinecite{carmeli2014spin} and Ref.~\onlinecite{gerster2012photocurrent}. In both works the same mutations of PSI were performed in order to covalently bind PSI to metal substrates (Ag and Au respectively). Ref.~\onlinecite{gerster2012photocurrent} found, for the bound PSI, the fastest excitation-relaxation cycle of around $15~\text{ns}$. For Ref.~\onlinecite{carmeli2014spin} the value should be on the same order of magnitude due to the large similarities between the two experiments. However, this value is still two orders of magnitude slower than the most ideal scenario that we have assumed. Therefore, the CISS-related spin signals in Ref.~\onlinecite{carmeli2014spin} was at least two orders of magnitude lower than the measured value. In fact, if we consider a realistic situation where PSI units do not form a fully-oriented and densely-packed layer on silver and $|\eta|<1$, the actual CISS signals should be even smaller.

Although other mechanisms may still be at play,\cite{nakayama2018molecular,cinchetti2017activating} they are not able to make up for the orders of magnitude of deviation. We thus conclude that the observed signals in Ref.~\onlinecite{carmeli2014spin} cannot be fully due to the light-induced spin injection from PSI, unless the very similar PSI conditions in Ref.~\onlinecite{carmeli2014spin} and Ref.~\onlinecite{gerster2012photocurrent} could lead to orders of magnitude of difference in PSI charge transfer time scales. This suggests that the magnetic-field dependence of the signals in Ref.~\onlinecite{carmeli2014spin} may predominantly originate from other effects. Some possible sources are discussed in Appendix~\ref{app:origin}.

Nevertheless, our analysis shows that an experimental approach as in Ref.~\onlinecite{carmeli2014spin} is in principle suited for confirming spin signals with CISS origin. It also provides insight in how one can optimize this type of experiments towards a system that would yield CISS spin signals with a higher magnitude. The most direct improvement can be obtained via a system that has higher values for $R_{sf}$ and $R_{eff}$ in Eqs.~(\ref{eqn:rsfag})-(\ref{eqn:reff-est}). A good example to consider is to use graphene as replacement for the silver layer. This should boost the spin signals by four orders of magnitude, since it would increase the value of $R_{eff}$ from $\sim$15~m$\Omega$ to a value of $\sim$0.5~k$\Omega$ (see Appendix~\ref{app:Reff} for details).

In summary, we introduced a two-current circuit-model approach to quantitatively assessing spintronic signals in hybrid devices which combine conventional electronic materials with (bio)organic molecules that are spin-active due to the CISS effect. As an example, we applied it to a case where the active layer has electrical contact only on one side, and we showed how the quantitative analysis can link the observed spin signals to charge excitation and relaxation times in the molecules. Our analysis showed that such devices can readily give spintronic signals that are strong enough for detection with current technologies. However, it also revealed that in the experiment of our case study (Ref.~\onlinecite{carmeli2014spin}), the observed signals must have had strong contributions from other effects. Future experimental work should aim at separating other signals from signals given by CISS, and our circuit-model approach assists in designing these experiments. We also recommend using devices with nonlocal geometries in order to separate charge and spin signals.\cite{yang2019spin,jedema2002electrical} In these geometries, the spin signals can also be quantitatively assessed using our circuit-model approach.

\section*{Acknowledgments}
The authors acknowledge the financial support from the Zernike Institute for Advanced Materials (ZIAM), and the Spinoza prize awarded to Prof. B. J. van Wees by the Nederlandse Organisatie voor Wetenschappelijk Onderzoek (NWO). We thank A.~Herrmann and P.~Gordiichuk for stimulating discussions.

\appendix

\section{Two-current model and derivation of $R_{sf}$ \label{app:2Cmodel}}
In this section we use a simple example to introduce the concept of two-current circuit models~\cite{mott1958theory,fert1968two} and to illustrate what can be experimentally detected. Along the way we derive $R_{sf}$ (Equation~\ref{eqn:rsfag}).

\begin{figure}[b!]

	\includegraphics[width=0.8\linewidth]{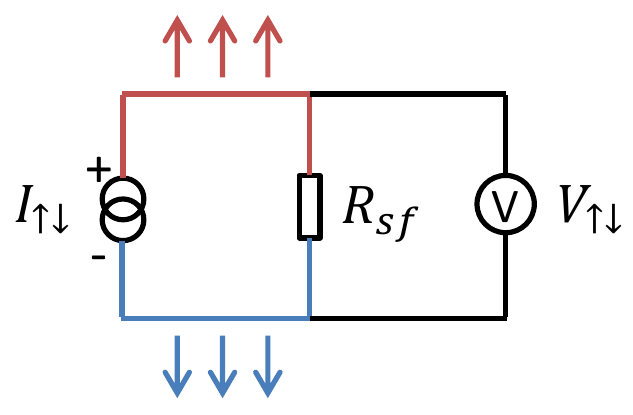}\\
	
	\caption{A circuit model considering the spin injection in a nonmagnetic conducting material. Spin-up and spin-down components are separated into red and blue channels. A pure spin current $I_{\uparrow \downarrow}$ is sourced between the two channels. Spin relaxation is modeled as a spin-flip resistance $R_{sf}$. The spin accumulation is measured as the signal $V_{\uparrow \downarrow}$ from a voltage meter that has fully spin-selective contacts.}
	\label{fig:silver}
\end{figure}

For describing spintronic signals, we use modeling where spin transport in conductors is described as two parallel channels each allowing only one type of spin (spin-up and spin-down channels, colored in red and blue respectively in Figure~\ref{fig:silver}).\cite{mott1958theory,fert1968two} This allows us to separate the total electrical current $I$ into spin-up and spin-down components: $I=I_\uparrow+I_\downarrow$. The difference between the two components is referred to as a spin current $I_{\uparrow \downarrow}$, with $I_{\uparrow \downarrow}=I_\uparrow-I_\downarrow$. A spin current injected into a non-magnetic material will result in a spin accumulation (chemical potential difference between the spin-up and spin-down channels) $\mu_{\uparrow \downarrow}= \mu_\uparrow-\mu_\downarrow $. Within the material spin accumulation decays exponentially over time due to spin relaxation mechanisms.\cite{vzutic2004spintronics} As an introduction to this type of modeling, we first show a simple case with a pure spin current in a nonmagnetic material, as shown in Figure~\ref{fig:silver}. A pure spin current means that the net charge current $I=I_\uparrow+I_\downarrow=0$. The spin relaxation is modeled as a pathway connecting the two channels, with a spin-flip resistance $R_{sf}$. The voltage difference between the two channels, as measured with fully spin-selective contacts, is therefore $V_{\uparrow\downarrow}=I_{\uparrow\downarrow} \cdot R_{sf}$.

Within the nonmagnetic material the steady-state spin accumulation is a balance between the spin injection due to $I_{\uparrow \downarrow}$ and the spin relaxation in the material. This is described as
\begin{equation}
0 = \frac{d \mu_{\uparrow \downarrow}}{d t} =
- \frac{ \mu_{\uparrow \downarrow}}{\tau_{sf}}
+ 2 \cdot \frac{I_{\uparrow \downarrow}}{e} \cdot \frac{1}{\nu_{3D} \cdot V_{rel} },
\end{equation}
where $\tau_{sf}$ is the spin-relaxation time in the material, $\nu_{3D}$ is the three-dimensional (3D) density of states (units of  $\text{eV}^{-1}\text{m}^{-3}$), and $V_{rel}$ is the relevant volume for the spin injection-relaxation balance in the material. The factor 2 arises from the fact that when one electron is transferred from the spin-down channel to the spin-up channel, the difference between the spin-up and spin-down population increases by two. The steady state solution for the measured voltage is
\begin{equation}\label{eqV-I-relation}
V_{\uparrow \downarrow} =\frac{\mu_{\uparrow \downarrow}}{e} = I_{\uparrow \downarrow} \cdot \frac{1}{V_{rel}} \cdot
\frac{1}{\nu_{3D}}  \cdot
\frac{2 \tau_{sf}}{e^2}.
\end{equation}

For further analysis we also consider the role of the spin relaxation length of the material, $\lambda_{sf}=\sqrt{D \; \tau_{sf}}$, where $D$ is the diffusion coefficient for electrons in the material. The Einstein relation gives $\sigma=e^2  \nu_{3D}  D $, where $\sigma$ is the conductivity of the material.\cite{jedema2002electrical} Consequently, Equation~(\ref{eqV-I-relation}) becomes
\begin{equation}
V_{\uparrow \downarrow}=2  \; I_{\uparrow \downarrow} \; \frac{\lambda_{sf}^2}{V_{rel}\;\sigma},
\end{equation}
and therefore:
\begin{equation}\label{eqn:rsf}
R_{sf} =\frac{V_{\uparrow \downarrow}}{I_{\uparrow \downarrow}}  =  2 \;  \frac{\lambda_{sf}^2}{V_{rel}\;\sigma}.
\end{equation}
We can see that the spin-flip resistance is completely determined by the properties of the material and the relevant volume concerned for each specific device.

Now we determine the relevant volume $V_{rel}$ for a particular device geometry: a thin layer of a nonmagnetic conducting material. The spin accumulation spreads out in a volume that is limited by either the spin relaxation length $\lambda_{sf}$, or the boundaries of the device, whichever is smaller. For the thin layer, we assume that the spin current is homogeneously injected from its top surface over a limited area, which is referred to as the relevant area $A_{rel}$. Spin accumulation then occurs in the thin layer within the area $A_{rel}$, as well as directly outside the boundaries of $A_{rel}$, up to a distance of $\sim$$\lambda_{sf}$. However, we consider here the situation where $A_{rel} \gg \lambda_{sf}^2$, and we can therefore neglect the spin accumulation outside $A_{rel}$. In the perpendicular direction we consider the case that the thickness of the layer $d < \lambda_{sf}$, which means that the spin-transport length is limited by the thickness of the layer rather than the spin relaxation length of the material. As a consequence, we have $V_{rel} = d \; A_{rel}$. Substituting this into Equation~(\ref{eqn:rsf}) gives
\begin{equation}\label{seqn:rsfag}
R_{sf} = 2 \;  \frac{\lambda_{sf}^2}{d \; A_{rel} \;\sigma}.
\end{equation}

When the thin layer (three-dimensional) is replaced by a truly two-dimensional material, such as graphene, the thickness of the material can no longer be defined. The material then has a two-dimensional density of states $\nu_{2D}$ (units of  $\text{eV}^{-1}\text{m}^{-2}$), and one should use the Einstein relation for the 2D conductivity $\sigma_{2D}=e^2  \nu_{2D}  D $. When assuming again $A_{rel} \gg \lambda_{sf}^2$, the spin-flip resistance for a 2D system is given as
\begin{equation}\label{eqn:rsf-2d}
R_{sf-2D}  =  2 \;  \frac{\lambda_{sf}^2}{A_{rel}\;\sigma_{2D}}.
\end{equation}

\section{Estimate for the value of $R_{eff}$ \label{app:Reff}}
In this section we estimate a value for the effective spinvalve resistance $R_{eff}$. We first focus on a value for the experimental work of Ref.~\onlinecite{carmeli2014spin}, and then on a similar system that has the silver layer replaced by graphene.

The tunneling resistance between silver and nickel was measured to be about $1~\text{k}\Omega$ in Ref.~\onlinecite{kumar2013device}, which used a device identical to that in Ref.~\onlinecite{carmeli2014spin}. The change of this resistance under magnetization reversal, as characterized by its tunneling magnetoresistance ($\text{TMR}=(R_{\downarrow}-R_{\uparrow})/R_{\uparrow}$), depends on the spin polarization of nickel $P_{Ni}$, and does not depend on the magnetization axis.~\cite{takanashi2015fundamentals,tedrow1971spin,moodera1995large} It follows $\text{TMR}=2 P_{Ni}/(1-P_{Ni})$, and takes a value of $\text{TMR} \approx 100 \%$ for the $P_{Ni} \approx 33\%$ value used in Ref.~\onlinecite{carmeli2014spin}. The actual TMR value may be lower than $100\%$ because of temperature and bias voltage, but should be on the same order of magnitude.~\cite{moodera1995large,lu1998bias} Moreover, taking the upper limit of TMR is consistent with us deriving the lower limit of $I_s$ and the upper limit of $\tau$. Therefore, we may assume $R_{\downarrow}=1~\text{k}\Omega$ and $R_{\uparrow}=0.5~\text{k}\Omega$. While this is an estimate, the value must be of the correct order of magnitude. Furthermore, later analysis will show that it is the spin-flip resistance of silver that governs the magnitude of the effective resistance $R_{eff}$.

To determine the spin-flip resistance of silver, we use the previously derived Equation~(\ref{seqn:rsfag}). For the device we discuss, the thickness of the silver layer $d= 50~\text{nm}$, the area of the junction $A_{rel}= 1~\mu\text{m} \times 1~\mu \text{m}$. For spin relaxation parameters we take reported values for a mesoscopic silver strip at room temperature $\lambda_{sf-Ag} \approx 150 ~\text{nm}$, and $\rho_{Ag}=1/\sigma_{Ag}\approx50 ~\text{n}\Omega\cdot\text{m}$.~\cite{godfrey2006spin} We point out that these parameters are not only affected by the material choice, but also by factors such as device geometry, fabrication techniques, and temperature.~\cite{bass2007spin} The values we chose were reported for a device that had geometries very close to that used in Ref.~\onlinecite{carmeli2014spin}, was fabricated with the same technique, and was measured at the same temperature. With these, we get the spin-flip resistance in our model $R_{sf-Ag}=45~\text{m}\Omega$.

Substituting $R_{sf-Ag}$, together with the assumed $R_{\uparrow}$, $R_{\downarrow}$ values in Equation~\ref{eqn:vdiff} gives an effective resistance
\begin{equation}
R_{eff} \approx 15 ~\text{m}\Omega .
\end{equation}
Note that $R_{eff}$ is fully determined by the properties of the Ag-AlO$_\text{x}$-Ni multilayer device, and estimating its value did not use any estimates or assumptions concerning PSI.

For the scenario where the silver layer is replaced by a graphene layer, we apply a similar analysis, while using Equation~(\ref{eqn:rsf-2d}) instead of Equation~(\ref{seqn:rsfag}). For graphene, typical material parameters are a square resistance of the order of 1~k$\Omega$,~\cite{guimaraes2014controlling,ingla201524} and a spin relaxation length of $\lambda_{sf} \approx 10~\mu\text{m}$.~\cite{tombros2007electronic,ingla201524} This gives $R_{sf-2D} \approx 1~\text{M}\Omega$, and $R_{eff} \approx 0.5~\text{k}\Omega$ for a device that is for other aspects identical to the device of Ref.~\onlinecite{carmeli2014spin}

\section{Analysis of compatible PSI excitation and relaxation times \label{app:times}}
In the main text we derived the $I_s$ values without using any information about PSI. Here we analyze what the values mean in terms of photo-excitation and relaxation times of individual PSI units. We first assume that $I_s$ is fully induced by the spin-selective electron transfer during photo-excitation and relaxation cycles in PSI. Then we examine the validity of this assumption by deriving (from $I_s$) the values of photo-excitation and relaxation times of individual PSI units. In the following discussion a few more assumptions are made. We carefully assume scenarios which consistently lead to the upper boundary of the photo-excitation-relaxation times. In the main text we showed that even this upper boundary is still too low to be realistic.

We write $I_s$ as a sum of the contributions from individual PSI units, 
\begin{equation}
I_s=\sum\limits_{n=1}^N i_{s,n}
\end{equation}
where $i_s$ is the spin current injected from each PSI unit into silver, the index $n$ runs over all individual PSI units, and $N$ is the number of PSI units within the relevant area (area of the junction) $A_{rel}$. We assume that all PSI units are oriented in the same direction, so that each of them contribute equally to the total current $I_s$. Therefore, we have $i_{s,n}\equiv i_s$, hence
\begin{equation}
I_s=i_s \cdot N = i_s \cdot \rho \cdot A_{rel}
\end{equation}
where $\rho$ is the number density, or coverage, of PSI. To estimate the coverage we need to take into consideration the size of PSI units. Isolated cyanobacterial PSI systems usually appear in trimers with typical diameters of around $30~\text{nm}$. This means three PSI units reside in an area of about $700~\text{nm}^2$, or for convenience, approximately a coverage of $\rho = 0.004~\text{nm}^{-2}$. Note that this is the highest possible coverage for a monolayer of PSI, since it corresponds to the entire silver surface being covered with a uniform, densely-packed PSI layer. We assume this maximum coverage for the entire junction area. We further assume that the total injected spin current $I_s$ is equally contributed by all the PSI units. This gives us an estimate of the lower boundary of $i_s$, the spin-current injection per PSI unit. For the \textit{up} orientation of PSI, we have $i_s\geqslant750~\text{pA}$. For the \textit{down} orientation this lower limit is $150~\text{pA}$.

Next, we analyze the magnitude of the charge current needed to produce this spin current via CISS effect. In our model, each PSI unit injects a spin current $i_s$ into silver, which is a fraction of the total spin current $i_{PSI}$ inside PSI. We have $i_s=\eta \cdot i_{PSI} $, with $-1\leqslant \eta \leqslant1$ being the fraction parameter. The value of $\eta$ depends on the spin-relaxation process inside PSI. In order to obtain a lower estimate of $i_{PSI}$, we assume $\eta=1$ (all the photo-induced spin current in PSI can be injected to the silver layer), hence $i_{PSI}=i_s$. This spin current, $i_{PSI}$, is again a fraction of the charge current $i$ induced by the continuous electron transfer during photo-excitation and relaxation cycles in a PSI unit. The conversion from a charge current into a spin current is due to the CISS effect and its efficiency is characterized by its polarization $P_{PSI}=i_{PSI}/i$. The CISS polarization of other chiral systems is reported to be about $50\%$,~\cite{gohler2011spin,carmeli2014spin,michaeli2016electron} so here we adopt the same value. Taking the above into account, we can derive the lower boundary of the charge current driven by photo-excitation and relaxation processes in a PSI unit: $i\geqslant1.5~\text{nA}$ for the \textit{up} orientation, and $300~\text{pA}$ for the \textit{down} orientation.

Finally, we translate this current into a value for the excitation-relaxation time $\tau$. Here, $\tau$ can be understood as the turn-over time, or the time interval between two consecutive photo-excitation processes from the same PSI unit. By assuming the intensity of the illumination is strong enough to drive all the PSI units in continuous excitation-relaxation cycles (saturated), we can write $i= -e/\tau$. A lower boundary of $i$ corresponds to an upper boundary of $\tau$. For the \textit{up} orientation of PSI, $i \geqslant 1.5~\text{nA}$ corresponds to $\tau \leqslant100~\text{ps}$. For the \textit{down} orientation the limit is $\tau \leqslant500~\text{ps}$.

\section{Possible origins of magnetic-field dependent signals in hybrid CISS devices \label{app:origin}}
There are other effects that can give rise to the magnetic-field-dependent signals in devices as used in Ref.~\onlinecite{carmeli2014spin}. One of these effects is the photo-response of silver. Any modification of the silver surface can change its work function. A work function as low as $1.8$~eV was reported for modified silver surfaces.~\cite{berglund1964photoemission_theory,berglund1964photoemission_experiment} It is therefore possible that the adsorbed PSI units and binder molecules modified the silver surface in a way that photoemission was allowed at the photon energies used in the experiment. This photoemission can be spin polarized due to the spin-orbit effect in silver and possible spin-dependent scattering at the surface.~\cite{kuch2001magnetic} Alternatively, the signals could also arise from a pure charge effect. Even without photoemission, the change of silver work function can lead to a voltage signal in the Ni-AlO$_\text{x}$-Ag capacitor. This voltage signal may depend on illumination and magnetic field, because the adsorbed PSI (which modifies the silver surface and thus the voltage signal) is highly photo-sensitive and contains large iron clusters that may respond to magnetic field. In such a scenario (where spin transport does not play a role), the orientation of PSI can only affect the magnitude but not the sign of the magnetic-field dependence. In fact, this is indeed the case if one considers the full signals reported in Figure~2A(ii) and Figure~2B(ii) of Ref.~\onlinecite{carmeli2014spin} instead of only their absolute values. In both figures, the measured signals can be separated into two parts: a nonzero background and a magnetic-field-dependent component that shows a step upon magnetic-field reversal. Figure~2B(ii) differs from Figure~2A(ii) by having an opposite sign for the background and a smaller step size upon magnetic-field reversal. The directions of the steps (i.e. the signs of the magnetic-field dependence) in both figures are the same: Both signals shift tens of nanovolts to less positive (more negative) values when reversing the magnetic field from \textit{down} to \textit{up} direction. The opposite signs for the background can be explained by the opposite orientations of PSI (just as how the PSI orientation affected the silver surface potential measured with a Kelvin probe), whereas the change of step size may be given by the change of position of the iron clusters with respect to the silver surface.


%

\end{document}